# Magnetic field-induced resistivity upturn and exceptional magneto-resistance in Weyl semimetal $TaSb_2$

Sudesh, P. Kumar and S. Patnaik

School of Physical Sciences, Jawaharlal Nehru University, New Delhi-110067, Delhi, India.

We study magneto-transport properties in single crystals of $TaSb_2$, which is a recently discovered topological semimetal. In the presence of magnetic field, the electrical resistivity shows onset of insulating behaviour followed by plateau at low temperature. Such resistivity plateau is generally assigned to topological surface states. $TaSb_2$ exhibits extremely high magneto-resistance (MR = $3.55 \times 10^4$ % at 2 K and 6 T) with non-saturating $B^n$ (n = 1.78) field dependence. We find that aspects of extremely large magneto resistance and resistivity plateau are well accounted by classical Kohler's scaling. Unambiguous evidence for anomalous Chiral transport is provided with observation of negative longitudinal magneto-resistance. Shubnikov-de Haas oscillations reveal two dominating frequencies, 201 T and 455 T. These aspects categorize $TaSb_2$ as a Type-II Weyl semimetal. At low temperature, the field dependence of Hall resistivity shows non-linear behaviour that indicates the presence of two types of charge carriers in consonance with reported electronic band structure. Analysis of Hall resistivity imply very high electron mobilities ($5.1 \times 10^4$ $cm^2 V^{-1} s^{-1}$ at 2K).

The nascent classification of materials in accordance with the topological states of quantum matter has yielded a radically new technological paradigm[1,2]. This relates to the discovery of several material systems belonging to topological insulators (TI) and topological semimetals (TSM) that exhibit extremely large magnetoresistance (XMR)[3–9]. Quite generally, the study of electrical resistance in the presence of external magnetic field renders deep insight into electronic transport mechanism[10] that collaterally leads to several technological applications such as in magnetic sensors, magnetic switches and magnetic storage devices. To bring in a prospective, the reported[6,11–19] magnetoresistance in TSMs could be two orders of magnitude higher than the giant magnetoresistance (GMR) or colossal magnetoresistance (CMR) observed in metallic thin films[20], perovskite manganites[21] or Cr-based chalcogenide spinels[22]. At the core of such exceptional magneto-resistive properties is the peculiar band structure of three-dimensional TIs and TSMs that yields conducting surface states. Recent studies reveal extremely large magnetoresistance (XMR) in Dirac semimetals (with linear band crossing at the Fermi level e.g. $Na_3Bi$[23] and $Cd_3As_2$[24]) that extends to Weyl semimetals like TaAs family[25,26] and compensated layered semimetals like $WTe_2$[27] and $MoTe_2$[28]. The XMR exhibited by these materials make them very interesting from the prospective of technological applications.

However, several recent works in a wide variety of topological materials have raised a fundamental question; whether the XMR can be explained by classical magneto-resistance theories without considering the topological aspects[15,29]. In terms of theories for TSMs, the associated symmetry principles restrict the quantum states to be robust against disorder due to time reversal symmetry invariance[30]. Experimentally this is manifested as a low temperature plateau in electrical resistivity. The plateau at low temperature is understood to have origin in conducting surface states that negate the insulating bulk behaviour, and are protected against backscattering due to time-reversal symmetry (TRS) invariance. The observation of large magneto-resistance in rare-earth monopnictides like R(Sb,Bi) (R = La, Y, Pr etc.)[18,19,31,32] and transition-metal dipnictides $TM_2$ (T = Nb, Ta and M = Sb, As)[33,34] have been analysed under such perspectives. In particular, $TM_2$ dipnictides are peculiar in the sense that at zero magnetic field they behave as weak topological insulators but with the application of external field, they can be classified as Type-II Weyl materials[35]. Several recent works have inferred that the plateau in these materials can also be explained by classical magnetoresistance theories without invoking topological surface states[15,18,31,33,36,37]. In this letter, we present a detailed study of magneto-transport behaviour in single crystalline $TaSb_2$. We observe XMR and metal-insulator like transition at low temperatures under external magnetic field. The resistivity plateau is observed at temperatures below 13 K. Our data on XMR and resistivity plateau is well accounted by Kohler's scaling. Further, observation of negative longitudinal magneto-resistance imply that TaSb$_2$ can be categorized as a Type-II Weyl semimetal. Moreover, clear Shubnikov-de Haas (SdH) oscillations are observed, at high magnetic fields and low temperatures revealing two major Fermi pockets.

Single crystals of TaSb$_2$ were synthesized by a two-step iodine vapour transport technique. In the first step, polycrystalline TaSb$_2$ was synthesized by solid-state reaction method. Stoichiometric amounts of Ta powder (3N, Alfa Aesar) and Sb shots (6N, Alfa Aesar) were reacted together by heating at 700°C for 3 days. The polycrystalline sample was then vacuum sealed with Iodine (50 mg/cm$^3$) in quartz ampoule and put in a tubular furnace with sample at 1000°C and temperature gradient of ~100°C across the sealed tube for one week. Shiny needle like single crystals were obtained in the size range 2-6 mm. The single crystallinity and crystal structure of the sample was analyzed using single crystal X-ray diffractometer (Bruker D8 Quest single crystal X-ray diffractometer). The structure of the sample was also studied using high resolution transmission electron microscope (HRTEM). Magnetotransport measurements in the temperature range 2 – 300 K and field range 0 – 6 T were performed in *Cryogenic* made cryogen free magnet (CFM) system. SdH oscillation measurements were performed using *Cryogenic* 14 T Physical Properties Measurement System (PPMS).

Single crystal X-ray diffraction data was collected using a microfocus anode (Mo) and a CMOS detector (PHOTON 100). The data analysis suggests a monoclinic (C$_{12/m1}$) structure with lattice parameters $a$ = 10.39(±0.03) Å, $b$ = 3.66(±0.01) Å, $c$ = 8.42(±0.03) Å, $\beta$ = 121.38(±13) and $V$ = 273(±2) Å$^3$ which is in accordance with the reported data for TaSb$_2$[17]. Inset in figure 1(a) shows the HRTEM image of the single crystal. The lattice fringes reveal the excellent crystalline nature of the single crystal. The inter-planar spacing is ~0.286 nm which corresponds to the orientation of (111) atomic planes of the monoclinic TaSb$_2$. Main panel of figure 1 shows the temperature dependent resistivity $\rho(T)$ of TaSb$_2$ sample in the absence of external field. The sample displays metallic nature. The resistivity decreases monotonically down to 2 K, with $\rho$(2 K) = 0.54 µΩ-cm, which indicates negligible defect scattering in the sample. The residual resistivity ratio ($RRR = \rho(300 K)/\rho(2 K)$) of the sample is observed to be ~260, implying high degree of crystallinity of the sample. At low temperatures, $\rho_{xx}(T)$ can be fitted with $\rho_{xx} = \rho_{xx}(0) + AT^n$ with n~2.56, where $\rho_{xx}(0)$ is the residual resistivity (= 0.538 µΩcm) and A = (4.29 × $10^{-10}$ ΩcmK$^{-n}$). The exponent value n is reflective of dominant scattering mechanism in the system with limiting values n = 2 for strong electron-electron scattering and n = 5 for conventional electron- phonon scattering processes. The observed value for TaSb$_2$ would imply dominance of limiting electron –electron scattering as was observed in NbP[38].

An extensively studied aspect of magneto-transport in TSM is the reported negative magnetoresistance when magnetic field is applied parallel to current direction.  Such observations are a common feature in several Weyl and Dirac semimetal such as TaAs, Cd$_3$As$_2$, and ZrSiS. Dirac semimetals are essentially gapless semiconductors with linear dispersion that become Weyl semimetals when Dirac point splits into two Weyl points due to either spatial inversion or time reversal symmetry breaking.  The chiral transport current between the Weyl points is not conserved leading to what is referred to as Adler-Bell-Jackiw (ABJ) anomaly[39], the experimental manifestation of which is negative magnetoresistance (for *B* ∥ I).  Figure 1(b) shows the transverse (*B* ⊥ I) magnetoresistance at several temperatures with current along *b*-axis and magnetic field along *c*-axis. At 2 K and 6 T, the sample shows extremely large magnetoresistance (MR = [[$\rho(B)$-$\rho(0)$] /$\rho(0)$] ×100 with $\rho(B)$ and $\rho(0)$ being electrical resistivity at applied B field and 0 field, respectively), with magnitude ~ 3.55×$10^4$ %, without any trace of saturation. With increase in temperature, MR decreases to 300% at 50 K, 6 T. Inset in figure 1(b) compares the field dependence of transverse and longitudinal magnetoresistance (LMR) at 2 K. In the longitudinal configuration, magnetic field and current are applied along the b-axis of the sample. We observe negative magnetoresistance above ~5.5 T below which a parabolic field dependence is seen. The possible reasons for the negative LMR could be (i) the magnetism in the sample that can be ruled out in TaSb$_2$, (ii) Improper contact geometry (non-uniform current) may also give rise to negative LMR[40]. Our specimen was needle –like and that eliminated such possibilities, and (iii) The emergence of Weyl points with application of magnetic field leading to Chiral anomaly in current transport.  In agreement with band structure of TaSb$_2$, we find this observation to be a clear signature of ABJ anomaly, which is observed in the form of longitudinal negative magnetoresistance. A recent theoretical report[35] confirms the possibility of hidden Weyl points in TaSb$_2$ that appear under applied magnetic field resulting in chiral anomaly.

In figure 2(a) we show Kohler's scaling with regard to constant temperature field scans. Kohler's rule gives a classical description of electronic motion that can provide insight into the MR behaviour in the sample. According to Kohler's scaling: MR = α(*B*/$\rho_0$ )$^m$ where α and m are sample dependent constants, $B$ is the applied field strength and $\rho_0$ is the zero field resistivity. Employing the Kohler's rule: MR = $\alpha(B/\rho_0)^m$, MR($B$) data are fitted for all temperatures against $\alpha(B/\rho_0)^{1.78}$ ($\alpha = 1.13 \times 10^4$). Inset (i) of figure 2(a) shows MR as a function of $B/\rho_0$ (at 2 K) from which $\alpha$ and m were determined. The collapse of MR data at all temperatures to a single line in the Kohler plot (main frame of figure 2(a)) implies that the sample shows similar power law for MR at all temperatures implying similar scattering mechanism is followed by the carriers at all temperatures. Inset of figure 2 (b) shows resistivity $\rho_{xx}(T)$ in applied perpendicular magnetic fields ranging from 0 to 6 T. We note that in the presence of magnetic field, above ~100 K, TaSb$_2$ shows metallic temperature dependence similar to the zero field $\rho_{xx}(T)$ behaviour. In the presence of magnetic field, the resistivity shows sharp upturn and drastic increase at temperatures below ~100 K. Moreover, below around 13 K, the resistivity starts to saturate, leading to a plateau-like behaviour in the $\rho_{xx}(T)$ curve. The temperature where the sample starts showing saturation remains unchanged at all applied magnetic fields. Such sharp upturn in resistivity below a particular temperature has been seen in several topological semimetals and associated XMR[31,39,41,42,7,43] has been reported. The microscopic explanation of such phenomena has been attempted with theories that include, a) magnetic field induced metal to insulator (MIT) transition, b) electron-hole compensation, and c) high mobility transport in metallic surface states of topological materials. However as recently reported for LaSb[44], the upturn in resistivity and its eventual saturation can also be ascribed to classical magnetoresistance theories involving Kohler scaling without invoking topological surface states. Kohler scaling can also be written as: $\rho_{xx}(T,B) = \rho_0 + \alpha B^m/(\rho_0)^{m-1}$. Since $\rho_0$ is the only temperature dependent term in this equation, the temperature variation of $\rho_{xx}(T,B)$ is mainly governed by $\rho_0$. Evidently, the second term ($\Delta\rho = \alpha B^m/(\rho_0)^{m-1}$) and $\rho_0$ have opposite dependence on temperature, leading to a minimum in total resistivity, $\rho_{xx}(T,B)$, at a particular temperature. The main panel of figure 2(b) shows the temperature dependent behaviour of $\rho_{xx}$(6 T), $\rho_0$ (= $\rho_{xx}$(0 T)) and $\Delta\rho$ (=$\rho_{xx}$(6 T) - $\rho_0$). The Kohler fitting to $\rho_{xx}$(6 T) is shown by dark blue line and the fitting parameters $\alpha$ and m are taken as $1.13 \times 10^4$ and 1.78 respectively. Evidently, the $\rho_{xx}$(6 T) in this figure fits well in the entire temperature range down from 2 K to 150 K that includes the plateau region as well. The correct crossover temperature of ~ 13 K is also verified from Kohler scaling. At low temperature, since $\rho_0$ is small and independent of temperature variation, it implies $\Delta\rho \gg \rho_0$ and $\rho_{xx}(T,B) \approx \Delta\rho \propto 1/\rho_0^{m-1}$. This signifies the presence of a plateau at low temperatures. Moreover, since both the resistivity upturn and plateau behaviour are explained by same Kohler's scaling, it implies that these peculiar behaviours are well explained by electronic transport across surface states. In summary, the low temperature emergence of plateau and XMR in TaSb$_2$ can be explained with the help of Kohler's scaling. In conjunction with the observation of ABJ anomaly, we conclude that Kohler's analysis is applicable to conducting states of Weyl semimetal and there is no apparent contradiction between Kohler scaling and surface transport in TSMs.

The Hall resistivity measurements were performed to investigate the carrier type, concentration and mobility. The field dependence of Hall resistivity $\rho_{xy}$ is shown in the inset (ii) of figure 2(a). The negative sign of $\rho_{xy}$ indicates possible dominance of electronic transport although, a large difference in mobilities can also give rise to same result. The Hall resistivity at 2 K is fitted with the two-band model[45] to evaluate the charge carrier concentration and mobility:

$$\rho_{xy} = \frac{B}{|e|} \frac{(n_h \mu_h^2 - n_e \mu_e^2) + (n_h - n_e)(\mu_h \mu_e)^2 B^2}{(n_h \mu_h + n_e \mu_e)^2 + (n_h - n_e)^2 (\mu_h \mu_e)^2 B^2}$$

here $n_e(n_h)$ and $\mu_e(\mu_h)$ are density and mobility of electrons(holes), respectively. The constraint: $\rho_{xx}(B = 0) = 1/e(n_e\mu_e + n_h\mu_h)$ is used to fit the Hall resistivity. The obtained values of electron and hole carrier concentrations are estimated to be ~$1.02 \times 10^{18}$ cm$^{-3}$ and $1.01 \times 10^{18}$ cm$^{-3}$, respectively, revealing TaSb$_2$ to be a compensated semimetal. From the fitting parameters, the obtained electron and hole mobilities are estimated as ~ $5.1 \times 10^4$ cm$^2$V$^{-1}$s$^{-1}$ and $1.36 \times 10^4$ cm$^2$V$^{-1}$s$^{-1}$, respectively.

Further, the magnetoresistance measurements show clear Shubnikov-de Haas (SdH) oscillations at low temperatures and high magnetic fields (see figure 3 (a)). Inset (i) shows zoomed data between 11 T - 13 T. The oscillation component (d$R$) of the MR is extracted by subtracting a higher order polynomial fit from the high field oscillation data. The oscillations can be identified from the plots of d$R$ Vs. $B^{-1}$, as shown in the inset (ii) of figure 3(a). In figure 3 (b) we show Fast Fourier transformation (FFT) of the data shown in inset (ii) of figure 3 (a). The oscillation frequencies are identified as 201 T (β) and 455 T (γ). From these SdH oscillation frequencies, the extremal cross-sectional area, $A_F$, of the Fermi surface can be extracted using the Onsager relation: $F = (\Phi_0/2\pi^2)A_F$. The frequency of $F = 201$ T corresponds to $A_F = 0.019$ Å$^2$ and $F = 455$ T corresponds to $A_F = 0.043$ Å$^2$. Also, we obtain Fermi wave vector, $\kappa_F$ (= $(A_F/\pi)^{1/2}$) = 0.078 Å$^{-1}$ and 0.117 Å$^{-1}$ corresponding to frequencies 201 T and 455 T, respectively. The FFT amplitude decreases with increasing temperature. In the inset of figure 3 (b) we show FFT amplitude with increasing temperature for β and γ peaks. From the temperature damping of SdH oscillation amplitude, the effective quasiparticle mass (m*) can be extracted. The value of m* is extracted from the fit of temperature dependence of FFT amplitude with Lifshitz-Kosevich (LK) equation: $\Delta\rho/\rho \propto AT/\sinh(AT)$, where $A = 2\pi^2\kappa_B m^*/\hbar eB$. The LK fit is shown in the inset of figure 3(b). The value of m* obtained from the LK fit for frequency 201 T is $0.17 m_e$ and that for frequency 455 T is $0.15 m_e$, where $m_e$ is the free electron mass. The Fermi velocity $v_F = \hbar\kappa_F/m^*$ is estimated to be $5.29\times10^5$ m/s and $7.94\times10^5$ m/s, respectively corresponding to frequencies 201 T and 455 T. The SdH oscillatory component was further analysed using the expression: $dR \propto \cos[2\pi(F/B-\gamma')]$[9], here $F$ is the frequency of oscillation and $\gamma'$ is the Onsager phase. From corresponding Landau fan diagram (not shown), the obtained $\gamma'$ was estimated to be zero confirming non-trivial Berry phase in $TaSb_2$[17].

To summarize, we present a magneto-transport study in the topological semimetal $TaSb_2$. At 2 K and 6 T a large transverse MR (= $3.55\times10^4$ %) is observed without any sign of saturation. The magnetic field induced turn-on behaviour and plateau like feature at low temperature are explained from the point of view of Kohler scaling. Significantly we find evidence for negative longitudinal magnetoresistance that signifies the presence of topological Weyl points. Non-trivial Berry phase is also indicated from analysis of Landau fan diagram. Thus, we demonstrate that the XMR and low temperature plateau explanation with the help of Kohler's scaling does not necessarily rule out the topological aspect of TSM. The high field SdH oscillations show two dominant frequencies at 201 T and 455 T. The Hall measurements confirm compensated semi-metallic behaviour with exceptionally high mobilities.

**Acknowledgements**

Sudesh and P. Kumar acknowledges DSK-PDF fellowship from UGC (Government of India) and JNU (New Delhi) fellowship, respectively, for financial support. Authors are thankful to AIRF (JNU) for access to the PPMS and TEM facilities. Low–temperature high magnetic field at JNU is supported under the FIST program of DST, Government of India. SP thanks SERB-DST for the project EMR/2016/003998/PHY. We thank Dr. Dinabandhu Das for advice on single crystal diffraction data analysis.

**Figure Captions**

**Figure 1:** (a) Temperature dependence of resistivity in zero applied magnetic field. Inset shows high resolution TEM image showing the planes (111). Solid line is a fit to the relation $\rho_{xx}$(T, B) = $\rho_{xx}$(0 T) + $AT^n$. (b) Field dependence of resistivity at various temperatures. Inset shows the field dependence of transverse ($B \perp I$) and longitudinal resistivity ($B \parallel I$) at 2 K.

**Figure 2:** (a) Inset (i) Kohler scaling of MR behaviour, MR = $\alpha(B/\rho_0)^m$. Inset shows the MR fitted with MR = $\alpha(B/\rho_0)^m$ with m = 1.78 at 2 K. Inset (ii) Field dependence of Hall resistivity at Temperatures 2 K, 50 K, 200 K and 300 K. (b) Temperature dependence of $\rho_{xx}$ (0T), $\rho_{xx}$(6T) and their difference $\Delta\rho$ are shown. Solid lines are fit to $\alpha 6^m/(\rho_0)^{m-1}$ and Kohler scaling. Inset in (b) shows temperature dependence of $\rho_{xx}$ at various magnetic fields.

**Figure 3:** Field dependence of resistivity at low temperatures (2 – 6 K) in the field range 0 – 13 T. Inset (i) shows the clear view of the SdH oscillations in the magnetic field range 11 – 13 T. Inset (ii) shows the oscillatory component (d$R$) of the resistivity at all temperatures (2 – 6 K) after subtracting the background. (b) FFT is plotted as a function of frequency at temperatures 2 – 6 K. Two dominant frequencies are obtained at 201 T and 455 T. Inset shows the FFT amplitude corresponding to frequencies 201 T and 455 T as a function of temperature. Solid lines show the fit of the data with Lifshitz-Kosevich formula.

Figure 1

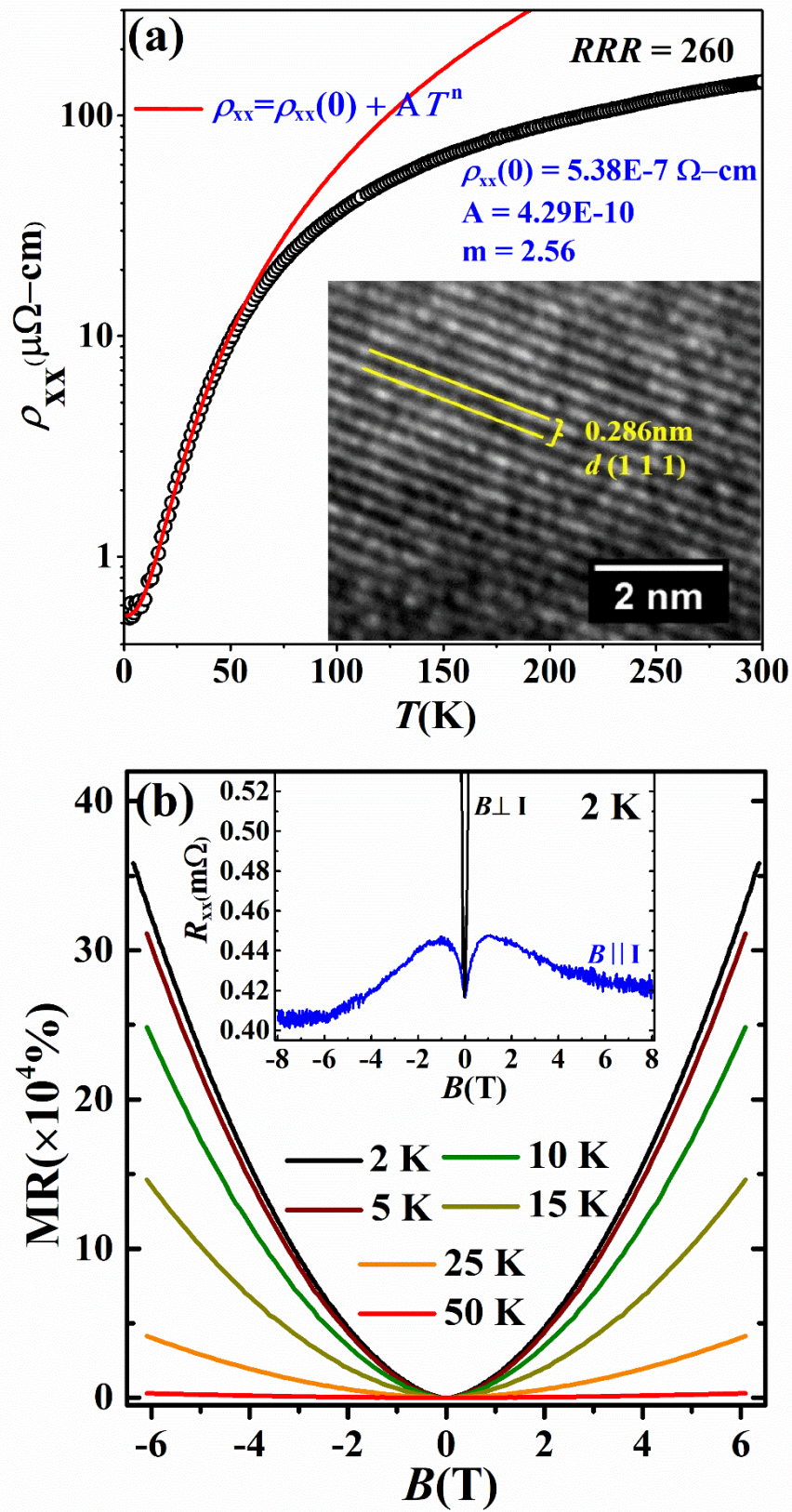

Figure 2

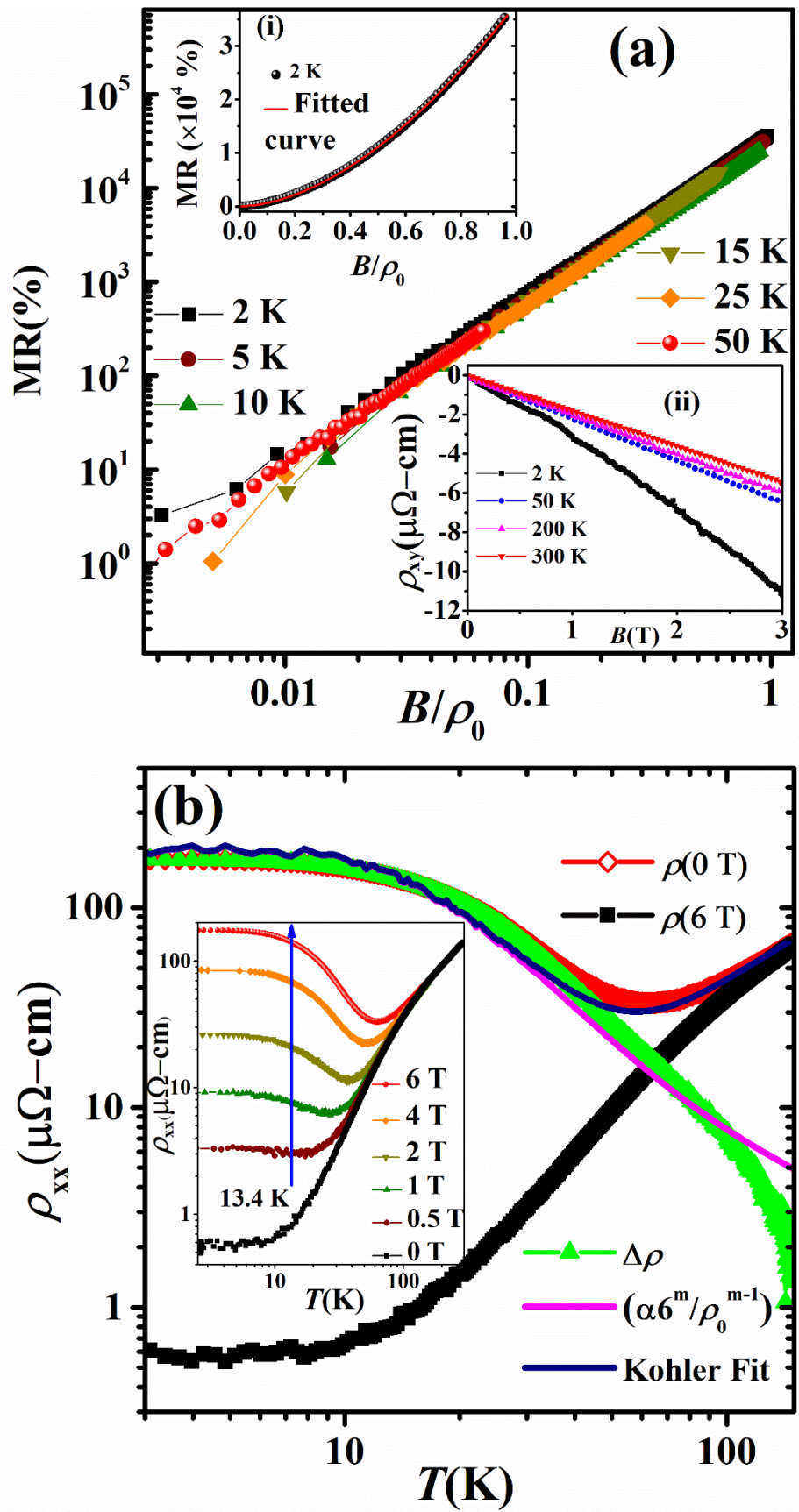

Figure 3

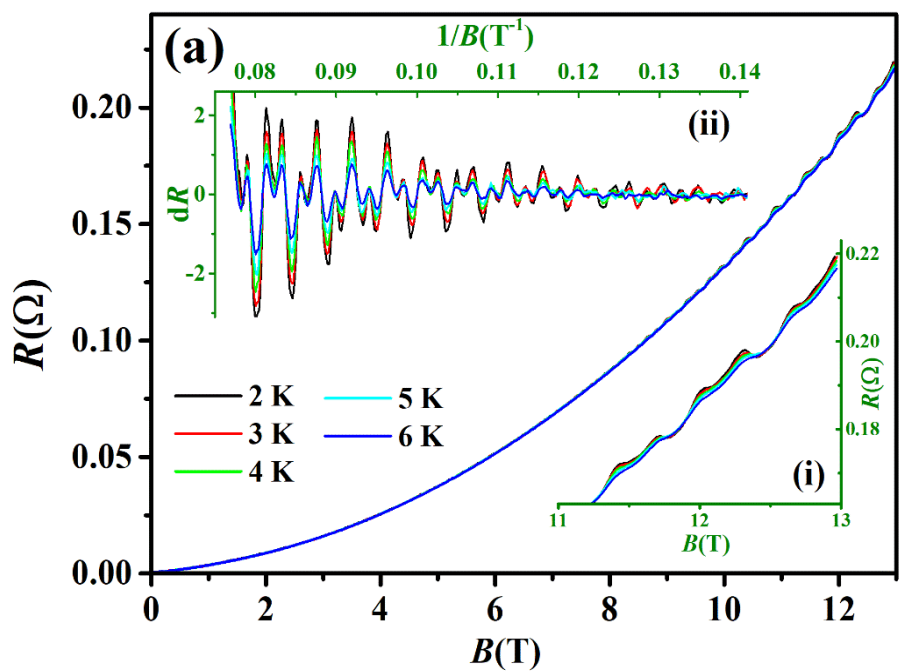

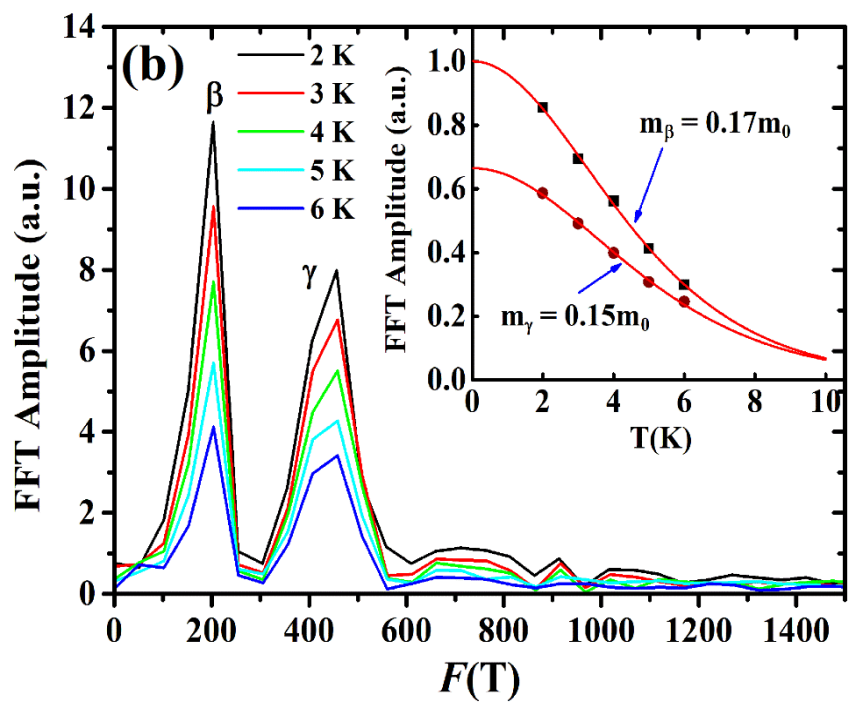